\documentclass[10pt, letterpaper]{llncs}
\title{SoK: Current State of Ethereum’s Enshrined Proposer Builder Separation}
\author{Maxwell Koegler \orcidlink{0009-0004-3166-6355}}

\institute{Lake Oswego Senior High School}

\usepackage{cite}
\usepackage{amsmath}
\usepackage{graphicx}
\usepackage{blindtext}
\usepackage{orcidlink}

\begin{document}

\maketitle
\begin{abstract}

    Initially introduced to Ethereum via Flashbots' MEV-boost, Proposer-Builder Separation allows proposers to auction off blockspace to a market of transaction orderers, known as builders. PBS is currently available to validators through the aforementioned MEV-boost, but its unregulated and relay-dependent nature has much of the Ethereum community calling for its enshrinement. Providing a protocol-integrated PBS marketspace and communication channel for payload outsourcing is termed PBS enshrinement. Although ePBS potentially introduces native MEV mitigation mechanisms and reduces validator operation costs, fears of multiparty collusion and chain stagnation are all too real. In addition to mitigating these potential drawbacks, PBS research pursues many tenets revered by Web3 enthusiasts, including but not limited to, censorship resistance, validator reward equity, and deflationary finance. The subsequent SoK will identify current PBS mechanisms, the need for enshrinement, additions to the ePBS upgrade, and the existing or potential on-chain socioeconomic implications of each.
    
\end{abstract}

\pagenumbering{arabic}
\setcounter{page}{1}

\section{Introduction}

    Currently, there are three main protocol-sponsored reward categories for honest Ethereum validators: proposal, attestation, and sync committee rewards \cite{penalties}. Proposal rewards are earned when a validator is selected to propose a block to the network; this is simply the reward for honest chain advancement. Attestation rewards are smaller consistent rewards issued when a validator honestly attests to the validity of a target or source block. Sync committee rewards are distributed to 512 randomly selected validators that inform light clients of block headers \cite{penalities2}. However, it is outside the bounds of protocol-sponsored rewards where the bulk of validator profits arise. For the sake of simplicity, protocol-sponsored rewards will be explained in more depth, but later it is revealed that although the issuance weight of proposition is much lower than that of attestation, proposition turns out to be much more lucrative when calculating for priority fees and transaction scrounging. 

	As presented in the Ethereum protocol documentation, the maximum protocol-sponsored reward, or ``base reward'', is determined by the following network-validator conditions:

\[
\text{\cite{penalties}}
\textbf{ }
\text{effective\_balance} \times \left( \frac{\text{base\_reward\_factor}}{\text{base\_rewards\_per\_epoch} \times \sqrt{\sum \text{active\_balance}}} \right)
\]

Simply put, the higher the effective balance of a validator and the lower the validator count, the higher the base fee \cite{penalties,penalities2}. When considering fee distribution, overall protocol issuance increases with validator count, but issuance per individual validator decreases — resisted dilution, if you will.

This base compensation is then further reduced depending on actualized network participation (determined by choice, resources, RANDAO, etc. \cite{penalities2, penalities3}). Fee distribution follows this criteria \cite{penalties}:

\begin{itemize}
    \item Attesting to the last justified block or source checkpoint $\to \approx 22\%$
    \item Attesting to the next proposed justified checkpoint $\to \approx 41\%$
    \item Attesting to the head of the chain on which to extend $\to \approx 22\%$
    \item Light client sync committee participation $\to \approx 3\%$
    \item Proposing the next block $\to \approx 12\%$
\end{itemize}

    By dividing up this reward system, it is revealed that attestation accounts for roughly 85\% of protocol-sponsored incentives, leaving a dismissible 3\% for sync committee participants and a rather minor 12\% for block proposition \cite{penalties}. Although this 12\% does contribute to the overall staker yield, It is not what generates the headline-making multi-Ether returns. The desire to be selected for block proposition arises not in the form of protocol incentives, but Maximal Extractable Value and priority fees. MEV is an exceptionally intricate topic, varying in execution, metagame, and downstream economic impacts; deservingly, it has countless papers dedicated to its complexity. MEV presents when a validator is selected to propose a block, and they leverage the unique opportunity to re-order, insert, and omit transactions in a manner that results in additional profits. 

    For example, when a DEX liquidity pool or AMM begins to misrepresent token valuation throughout the course of the block, a proposer can insert a personal transaction to exploit this distortion. By trading against this ``warped'' liquidity curve, they gain a favorable conversion rate and can later trade through a balanced AMM/LP to secure a small profit \cite{LPcurves}; this particular MEV strategy is one of many ever growing in complexity and efficiency. Because MEV is not issued by the protocol, all extracted value comes from the pockets of genuine users and liquidity providers, making it widely considered ``toxic'' or harmful. 
    
	Priority fees on the other hand, or proposer ``tips,'' incentivize earlier transaction incorporation by effectively bribing proposers to be included. This direct payment from user to proposer is also excluded from the previous reward categories as it is not financed by the protocol. Typically, this fee is conscientiously added to transactions for faster execution, therefore making it an intentional fee market, and not one needing intervention beyond improving distribution equity. Between increasing network demand and limited blockspace, this expense is almost always a component of gas fees \cite{penalties, penalities2, penalities3}.
    
	Priority fees and MEV vary in magnitude depending on network congestion, market conditions, and user activities. Because no individual could induce any of these under reasonable circumstances, this condition distribution is widely considered ``fair,'' beyond circumstantial inequalities. Conversely, due to the wide variety of validator hardware, not every proposer can equally identify and exploit extractable MEV opportunities; this is widely considered inequitable and in need of correction, as validator's resources should not dictate reward capacity unless otherwise providing more service to the network. Furthermore, potential reward extremity makes MEV suppression or redirection a hot topic in the greater Ethereum community. Proper MEV management stands to not only protect genuine users, but also may act as an important step towards validator equality and a deflationary asset model.

\section{Current Approaches}

    As mentioned above, there are two constraints to MEV extraction — these include both the conditions and means to extract. In essence, LP distribution curves, mempool contents, and user activities determine potentially extractable value, then, depending on whether the proposer has the resources and computational capacity to identify these opportunities, their actualized extractable value is determined. Because potentially extractable value is relatively equitable (since proposition conditions are random), the main equality vector is actualized extractable value \cite{ViablePBS}. In an attempt to rebalance proposer rewards, means of outsourcing block generation has come about to provide weaker validators with the ability to compete for available MEV. Currently, this outsourcing is primarily facilitated by Flashbots' MEV-boost, with upwards of 90\% of validators leveraging it regularly \cite{mevprominence}.

\begin{figure}[ht]
  \centering
  \includegraphics[width=0.8\textwidth]{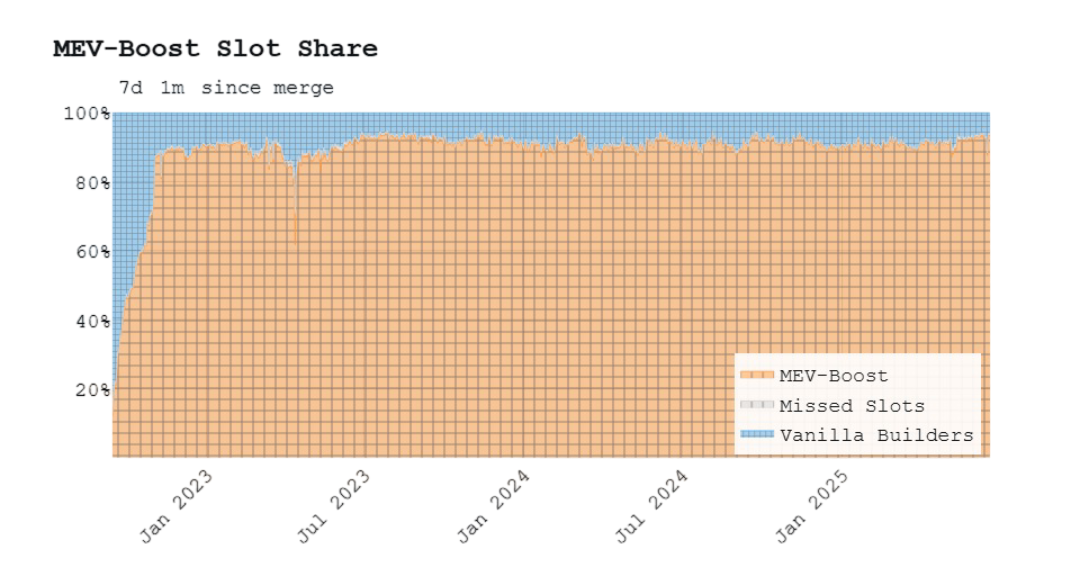}
  \caption{MEV-boost slot prominence since the merge (light orange) \cite{mevprominence}.}
  \label{fig2}
\end{figure}

	MEV-boost is a middleware addition to a validator's consensus client \cite{orderflowfig}. This open-source project essentially aggregates trusted relays and their subsequent builder markets to provide validators with a trusted communication channel for exchanging block contents and payments with builders. Builders are separate entities that specialize in maximizing slot value through transaction ordering and block structuring \cite{builderspecs}. Aside from merely providing this communication channel and marketspace, MEV-boost also leverages Flashbots' private orderflow to increase MEV capacity and availability, effectively increasing both potential and actualized extractable value for proposers. 

	The complete journey of an MEV-boost transaction begins with its submission. This can be in the form of a public mempool submission or a submission to an MEV-share affiliated RPC endpoint \cite{orderflowfig}. MEV-share is also an open-source Flahsbots project that enables the bypassing of Ethereum public mempools by submitting transactions directly to builders, and later MEV-boost leveraging proposers \cite{MEV-share}. When submitted to a public Ethereum mempool, a user exposes themselves to toxic MEV executed by non-validating actors. Although MEV-share transactions are still submitted to builders for MEV extraction, MEV-share participants receive a 90\% extraction kickback and do not expose their transactions to the public prior to execution. Once a user submits their transaction, either to a public or private mempool, their transaction may be chosen by searchers to bundle into an MEV bundle. These MEV bundles contain extractable MEV instances and are submitted directly to MEV-boost builders; this lightens their computational load by pre-compiling extractable transaction instances in exchange for a fee \cite{orderflowfig, MEV-share}. Regardless of whether a block is composed of searcher bundles, public mempool transactions, private orderflow transactions, or a combination of each, builders eventually have an array of transactions or bundles of which to compose a block.

\begin{figure}[ht]
  \centering
  \includegraphics[width=0.8\textwidth]{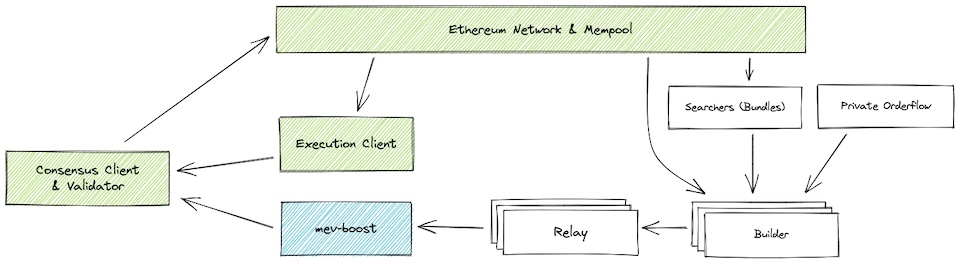}
  \caption{MEV-boost orderflow \cite{orderflowfig}. }
  \label{fig1}
\end{figure}

    With valid and profitable transactions available to choose from, a builder begins constructing their block. To clarify, most builders have access to the same key transactions unless tapped into some exclusive transaction source or a broader mempool radar. Builders must optimally order transactions not only to increase personal profits but also to remain competitive in upcoming bidding. Assuming that a builder can maximize the value of a slot through transaction ordering, they effectively increase the value of the blockspace and the minimum value threshold for inclusion (especially valuable during congestion events). Raising the economic efficiency of transaction inclusion proportionately decreases network congestion and node hardware demand, while increasing throughput via encouraged transaction efficiency. If a builder produces a more optimal block than others and can afford a higher bid, it will be selected by the proposer and the payload is exchanged; through this process, builders can assume the computational burden of ordering extractable transactions and the proposer can, for a fee, participate in MEV/mass tipping events with minimal hardware \cite{builderspecs,orderflowfig}. 
    
	Although this may sound great, there are many nuances to the MEV-boost procedure, especially considering the fact that it is not operated by the core protocol. Starting with relays; briefly mentioned earlier, relays facilitate the communication between builders and proposers, effectively protecting the builder's payload and the proposer's slot. The main issue with a third party relay is that it must be trusted by both parties, a difficult feat to achieve when hundreds of Ether are at stake. Mutual distrust between builders and proposers makes relay operation difficult as both parties are financially incentivized to steal block contents or deliver improper payments \cite{constraints,ViablePBS,orderflowfig, relayfundamentals}. The Flashbots consideration page also suggests that relays have the capacity to submit faulty bids without any financial penalty, an unfortunate consequence of no minimum stake and limited relay accountability \cite{considerations}. Despite this malicious potential, relays operate on a reputation system, in which validators can identify malicious activity through circuit breakers and relay monitors, informing others of relay misbehavior, and ideally removing its usage from general circulation \cite{relayfundamentals}.
    
	Despite the damage a dishonest relay stands to cause, its relevance depends solely on its performance for the proposer. This means that the only incentive to betray their proposer would be a significant MEV spike that the relay could extract \cite{relayfundamentals}. If identified by the dishonest relay, it could structure its block, bypass bidding rounds, submit a fraudulent bid, claim the MEV, and leave both its reputation and the proposer's trust in the past. Assuming other proposers become aware of this malicious activity, relay monitors and circuit breakers would immediately trip, and the fraudulent relay would lose all future credibility. Most notable relays are run by Flashbots themselves (a notable point of centralization), so assuming they aren't willing to sacrifice their entire enterprise for one extractable MEV instance, it's safe to say this fraudulence risk is minimal for most most relays \cite{relayfundamentals}.
    
	With off-chain systems, it is common for accountability to slip and corruption to appear on either end of a communication channel; faulty bids, hidden MEV, and external builder-relay relationships may skew the builder market, discourage open competition, and centralize block creation. Unfortunately, this technicality is not even a requirement for builder corruption; whether by relay priority, access to exclusive orderflow, or just a bidding reserve, if a builder ever gets a leg up on others, they can easily monopolize the block creation market \cite{considerations, constraints}. Enforced bid selection usually deters loyalty, but any builder with a bidding reserve only needs to outbid the second wealthiest bidder to claim the entirety of a blocks value. This dynamic often leads to cyclical dominance, where a builder repeatedly outbids competitors, extracts MEV, and expands their bidding reservoir uncontested. Without competition, potentially due to a variety of reasons, a few main builders emerge and an additional centralization point arises. Although builder advantage is likely to arise in ePBS as well, developed actor accountability may play an influential role in the reduction of these patterns. 
    
	It is important to note that much of this theoretical corruption is not the result of MEV-boost itself, rather the result of an unfortunately convenient foothold. Most, if not all of the aforementioned corruption, collusion, and network deterioration is the result of greedy MEV actors aiming to control the proposition market space \cite{considerations}. A defective relay or builder could easily wreak havoc on a clueless validator or an oblivious network; censorship, centralization, missed slots, and major chain stagnation are of the many risks associated with relying on third-party mediation. Although not all these struggles will be addressed in one upgrade, many can be mitigated. Before ePBS is discussed, Commit-Boost an emerging MEV-boost-compliant confirmation manager — must be briefly addressed. 

    \subsection{Commit-Boost and Preconfirmations}

    As a vibrant and efficient economy does, the myriad of available economic opportunities is always being exploited in new and creative ways. This efficiency, however critical it may be to the financial integrity of the network, does additionally contribute to fragmentation and excessive client additions. One such client addition is confirmation clients: a client sidecar that enables the requesting of future blockspace (eg. ``I want a validator to include this transaction if selected to propose for this slot'' aka preconfs). With so many additions required to maximize a validators viability in terms of commitment opportunity, home operators may experience pressure to \textit{join} a blockspace negotiation team, while other commercial validators may experience vendor lock-in, with migration costs too high to effectively upgrade their client(s). Additionally, with so many MEV-Boost-like blockspace negotiation clients, threat vectors, network reliance, and general turbidity could easily arise from such stiff client-solution relations. To solve this unwanted ``client-solution turgidity'' and network fragmentation, the Ethereum developer community has created Commit-Boost \cite{commit-boost}, a unifying yet modular Ethereum sidecar that standardizes proposer commitment opportunity while maintaining MEV-boost compatibility. So long as Commit-boost maintains an active and responsible client base, users can potentially speculate on future blockspace value (long-dated blockspace futures) or request (preconfirm) payload inclusion within a certain time range (preconfs); by unifying commitment logic under one client solution, operators can easily manage these potentially advanced preconf requests \cite{commit-boost}. 

\section{Enshrinement}

    The enshrinement of Proposer-Builder-Separation is commonly deemed ``Stage 3'' of PBS, after conceptualization and off-chain implementation. This improvement is developed in draft EIP-7732 \cite{7732} and aims to decentralize the procedure explained in the preceding chapter. By building PBS options into the core protocol, the approximately 90\% of validators currently dependent on MEV-boost \cite{mevprominence} will no longer have to trust independent relays or worry about builder-relay corruption. The five main motivations for this EIP are listed below \cite{7732}:

\begin{itemize}
    \item Trustless exchange between proposers and builders, ensuring proposers are paid and the builder's payload becomes the chain head
    \item Reduced computational demand on validators and the enabling of lighter validating clients
    \item Faster network propagation without full execution payload publishing
    \item Gives validators time to attest, strengthening the fork choice in case of invalid builder payloads
     \item Removes centralization risks and middleware dependence
\end{itemize}

	As discussed earlier, trust-dependent relays pose quite a challenge to network integrity and builder-proposer relationships. From potential collusion to invalid bids, it is clear that independent relays need to be addressed. EIP-7732 proposes that relays be effectively replaced by an on-chain auctioning system in which communications would be achieved directly through the consensus layer. Not only would this prevent relay favor and ensure bidding validity, but it would also guarantee a fair trade between proposers and builders without the need for third-party mediation \cite{7732}. Relay enshrinement is achieved by builders including their payload via a block header in the next beacon block. As each header reflects the associated bid, a proposer can monitor these headers and choose which it wishes to propagate according to its associated bid. Once the header is selected, the bid from the block header is executed, the builder's MEV is extracted, and the builder reveals their execution payload \cite{7732}. 
	
    The main concerns regarding off-chain PBS implementations are trusted relay dependence (which has now been addressed), and actor accountability (which will now be discussed). It is suggested by the EIP-7732 draft\cite{7732}, that a Payload Timeliness Committee (PTC) be assigned to oversee builder delivery, payload integrity, and the timeliness of the reveal. This committee would penalize any builder who did not promptly or honestly reveal their promised block through the boosting of the parent block's fork choice weight. This boost would equate to about 40\% of the beacon committee and would effectively count against the block's adoption as head. By contrast, blocks submitted in an honest and timely manner receive a 40\% boost supporting its selection as chain head \cite{7732}. Although this internally manages chain adoption of compiled blocks, managing a proposer's unconditional payment is another question.

    First, it needs to be understood when a proposer payment is owed. Naturally, if a proposer fails to select a beacon block or its associated execution payload for a slot, it never asked a builder to produce a block and is therefore liable for the missed slot. In other words, no one owes a proposer for its own shortcoming. This is referred to as a skipped slot by the EIP-7732 draft \cite{7732}. Alternatively, if a proposer includes a beacon block and therefore commits to a builders execution payload, but said payload is never revealed by the builder, the builder is liable. This is referred to as an empty slot by the EIP documentation \cite{7732}. If liability can now be concretely determined, the next step is enforcing accountability and ensuring proper compensation. Because the execution payload is revealed after proposer commitment, proposer-originating betrayal is technically impossible, leaving builder accountability as the only factor in need of addressing. This is achieved in the form of builder stakes. As pointed out by EIP-7732, builders now effectively act as a new leg of validation, earning their role a new staking requirement. By enforcing builder stake, or at least collecting evidence of a sufficient bidding reservoir, proposer payment will be issued regardless of a builder's performance \cite{7732}. This guarantees a proposer's payment and a builder's accountability, but further perpetuates the reservoir advantage mentioned in the preceding chapter. 
    
	As validators now have the option to outsource block development, their computational requirements are lowered and they are given more time to attest, potentially strengthening the fork choice rule in the case of an invalid builder payload \cite{7732}. Lighter clients may also better from less computational demand on block formulation. Additionally, network speed will not be dramatically affected as builder headers are the only broadcasts prior to post-selection payloads. 
    
	Moving the entire PBS procedure on-chain resolves a number of our previous concerns, including relay dishonesty, middleware dependence, and builder accountability, yet some negative characteristics still remain. Even with an enshrined PBS system, there remains a risk that a few well-resourced builders could dominate the bidding market and monopolize block production. So long as builders can subsidize their bidding with external funding (e.g. a reserve) there will always be the potential for resourced builders to outbid the competition to the point where it is no longer computationally efficient to compete. Eventually, every financially concerned builder will halt their operations in light of wasted computational resources, and dominant builders will be able to slacken bids. This assumes that eventually, all builders will give up after insistent overbidding by opposing builders; however, some may keep going, eventually draining the bidding reserves of the opposing builder and reinstating builder competition. 

	Despite relays being removed from the equation, it is still very possible for builders to collude, bids to drop, and transactions to be censored, all the more supported by a higher barrier-to-entry instated by builder staking. On the contrary, the same is said for proposers; if they collude, manage to secure 20\% of the total stake, and land multiple consecutive blocks they could reorg a builder payload and redraw the bid. Of course, this all is theoretical and highly improbable, as colluding parties would need nearly 6.7M Ether (a point at which the yield would be negligible) \cite{constraints, 7732}. One final note, exclusive orderflows still exist, but now aren't communalized among MEV-boost builders; unless Flashbots' MEV-share and alike exclusive orderflows are dissolved or made public, some builders may have disproportionate access to transactions, potentially leading to builder dominance and centralization. 
    
	ePBS stands to resolve many of the bottlenecks, centralization points, and corruption risks associated with MEV-boost, but, nonetheless, there is still much more to work on. Although EIP-7732 enshrines the auctioning process of PBS, questions of MEV and censorship still stand. As of now, the official ePBS EIP supports what has been mentioned, but does not address selective inclusion by builders or persistent MEV dynamics. If implemented in its current state, MEV will still stand to affect users, liquidity providers, and proposer equality (in the form of potentially extractable value). 

\section{ePBS additions}

Although the EIP-7732 draft represents a significant step towards enshrined Proposer-Builder-Seperation and validator equality in Ethereum, it remains a work in progress, with ongoing development and community discussions shaping its finalization. EIP-7732, the formalization of ePBS, stands to not only reshape how validators propose blocks, but may also be one of the biggest steps toward micro-validation and hyperlight clients. The greatest achievement of ePBS is perhaps, its ability to pave the way for MEV management mechanisms and stronger user protection schemes. Below are the most recognized redistributive primitives in the field.

\subsection{Committee-driven MEV smoothing}

	Francesco D'Amato's \textit{``Committee-driven MEV smoothing''} \cite{MEV-smoothing} highlights the capacity for attestation committees to be used as MEV distribution pools that effectively `smooth out' large MEV spikes. Communalizing MEV rewards reduces variance and increases reward equity for validators. It was proposed that attestation committees encourage ``cooperative equilibrium'' through the enforcement of highest-bid selection. Every time a validator within any given attestation committee is selected to propose a block, their MEV and priority fee total is distributed between committee participants. This way, if a validator ever receives extreme MEV/priority fees, this can be `smoothed out' among committee members, creating a more equitable and consistent validation incentive scheme. 

\subsection{Spam resistant block creator selection via burn auction}

The problem with the aforementioned MEV-smoothing and its associated distribution method, is that users never recover the value extracted from them; it is still taken from users and circulated among validators. This can be solved by MEV burning. Barry Whitehat of the Ethereum Foundation notes in his summary of burn auctions, \textit{``Spam resistant block creator selection via burn auction''} \cite{SPAM} that proposers are always looking for builders to prove their blocks are maximally productive. This in PBS and ePBS, is demonstrated via direct bids to proposers, but in this burn auction proposal, it is noted that this proof can be in the form of burn commitments. Rather than promising payment to a validator, builders can commit to burning a certain amount of Ether. By doing so, the proposer can identify the block with the most efficient MEV extraction simply by how much the builder is capable or willing to burn for the inclusion of their block. The main benefit of burning this MEV is the refunding of extracted value back to the pockets from which it came. Considering that burning Ether reduces its circulation and therefore increases its value, all Eth-holding users will benefit from this public good. Although asset burning does not give the Ether directly back to the users of which it came, they are getting some diluted public good from the perpetual burning of Eth.

One issue arising from the use of this mechanism is that builders are incentivized to win these auctions and keep the remaining, unburnt MEV, while the proposers have no specialized incentive to propagate more MEV-burning blocks. Unless otherwise forced to choose a high-burning block by some enforcement committee, the validator would have a better chance at getting MEV rewards by colluding with builders privately and choosing a low-burning block to secure the remaining MEV. Naturally, enforcement would be assumed, and this prediction nullified. If introduced in tandem with the EIP-7732, proposers would be strictly rewarded with protocol-sponsored incentives, and the downstream public good of burning, while staked builders have a chance at small MEV margins after burning their bid. Accordingly, this ePBS addition effectively distributes MEV volatility among a proposer, a builder, and all Ether holders. To clarify, ePBS participation (as of EIP-7732) is optional \cite{7732}, yet some of these additions make it beneficial to avoid block outsourcing and instead independently construct blocks, regardless of if your a weaker validator or not. In order to make these burn auctions effective and equitable, ePBS enforcement should be considered. 

\subsection{MEV burn—a simple design}

Another attempt at managing MEV is proposed in Justin Drake's \textit{``MEV burn—a simple design''} \cite{MEV-burn}, where builders are required to burn a certain subjective base fee when their payload is selected, whether or not the payload was accurately and honestly delivered. This base burn is somewhere below a builder's effective balance, minus their payload tip. This is analogous to burn auctions with a variable block priority fee. This also mimics the raw transaction structure we see today; nearly as if extrapolating EIP-1559 \cite{1559} to block formulation. Once bidding rounds conclude, a dedicated attestation committee monitors the highest observed base fee and informs the proposer that they must choose a bid above this value. Considering that this base fee was the highest at selection, the only available bids to choose from are ones greater than this floor, sent after the committee session \cite{MEV-burn}. Knowing this, a proposer is most likely to choose, from this restricted array, the highest payload tip, simply out of self-interest. In this model, builder collusion would nearly be the only way to reduce the base fee. If all the builders identify the committee meeting time and wait until afterward to bid, they can submit floor zero bids and allocate the rest of their extraction to tips. However, this would not benefit the builder much, as most of their profit would then just be dwindled down again by more aggressive tipping. Unless otherwise refunded by proposer kickbacks, builders are financially discouraged from this fee avoidance, as they neither receive burn benefits nor higher fee margins. Because of this, it can be assumed that the risk of collusion between all bidders and the proposer is negligible. 

\vspace{\baselineskip} 

These ePBS additions could leverage this valuable opportunity to make a lasting cryptoeconomic impact on the MEV ecosystem. Mitigating centralization risks, instating validation equality, and protecting everyday users are, of the many outcomes resulting from adequate EIP-7732 development. There are many more proposed MEV mitigation mechanisms that expand upon the ePBS assumption, but listing them all would be redundant. The provided examples give a window of insight into the three main categories of MEV management, distribution (validator reward spreading), mitigation (pure-burn auctions), and a mix of both (reward distribution post-burn). As the Ethereum research community is exhaustive, more branches of MEV mitigation mechanics will surely emerge and solve the shortcomings of previous infrastructure.

\section{Conclusion and acknowledgments}

Countless frontline Ethereum researchers are hard at work discussing and solving the many complications of Proposer-Builder-Separation. So far, the need to enshrine current PBS procedures defined in Flashbots' MEV-boost is apparent, but the path to ePBS may not be. The preceding paper explored the following:

\begin{itemize}
    \item The desire for general PBS
	\begin{itemize}
		\item Inequitable MEV distribution among validators
		\item Sub-optimal economic extraction
	\end{itemize}
    \item MEV-boost, the predominant PBS facilitator
	\begin{itemize}
		\item Standard orderflow
		\item Relay trust dependence
		\item Corruption and collusion

		\item Participant accountability

		\item Extortion and bidding profiles

            \item Commit-boost

	\end{itemize}
	\item PBS Enshrinement
		\begin{itemize}
			\item EIP implementation
			\item Relay dissolution 
			\item Payload transmission methods
			\item Participant accountability (solved)
		\end{itemize}
	\item ePBS additions
		\begin{itemize}
			\item MEV committee-based distribution
			\item Burn auction implementation
			\item Subjective burning
		\end{itemize}

\end{itemize}

Alongside validator reward equity and overall network longevity, determining what is acceptable to claim as a proposition reward, what must be done to restrict inappropriate issuance, and what can be accomplished to further develop user protection schemes are all hot topics of this seemingly eternal discussion. From here I hope to inspire others to dive into this labyrinth of a topic, explore the intricacies of the Ethereum protocol, and leverage the countless resources that the Ethereum community has worked so hard to make available. 

To every researcher and Web3 artist cited in this paper, I thank you for your priceless insights and the critical role you play in the Ethereum research community; this SoK would be impossible without your hard work. Additional thanks to Mr. Michael Hall and Mr. Adam Dennis for their invaluable insight regarding the contents of this paper.

\bibliographystyle{plain}

\bibliography{references}

\end{document}